# Modeling Optical Fiber Space Division Multiplexed Quantum Key Distribution Systems

MARIO UREÑA,[1] IVANA GASULLA,[1] FRANCISCO JAVIER FRAILE,[2] AND JOSE CAPMANY[1,*]

[1]*ITEAM Research Institute, Universitat Politècnica de València, Camino de Vera, 46022 Valencia, Spain*
[2]*Dept. Teoría de la Señal y Comunicaciones, Universidad de Vigo, ETS Ingenieros de Telecomunicación, Campus Universitario, E-36310 Vigo, Spain*
*\*jcapmany@iteam.upv.es*

**Abstract:** We report a model to evaluate the performance of multiple quantum key distribution (QKD) channel transmission using spatial division multiplexing (SDM) in multicore (MCF) and few-mode fibers (FMF). The model is then used to analyze the feasibility of QKD transmission in 7-core MCFs for two scenarios of practical interest. First for transmission of only QKD channels, the second for simultaneous transmission of QKD and classical channels. In the first case, standard homogeneous MCFs enable transmission distances per core compatible with transmission parameters (distance and net key rate) very close to those of single core singlemode fibers. For the second case, heterogeneous MCFs must be employed to make this option feasible.

## 1. Introduction

Quantum key distribution (QKD) provides an intrinsically secure way to distribute such secret keys between remote parties [1-3]. The secrecy of the keys distributed by QKD is verifiable [4], since it relies upon quantum mechanical principles featuring resilience against an eavesdropper. Early QKD experiments focused upon the feasibility of the technology, starting from proof-of-principle laboratory experiments. Subsequent developments in the field dramatically improved the security, performance, accessibility and reliability of the QKD technology. Its security can be rigorously proven even when implemented with practical components only, such as attenuated lasers [5]. The secure bit rate has been increased by three orders of magnitude to 1 Mb/s over 50 km fiber thanks to the development of efficient QKD protocol [6] and high-speed single photon detectors [7].

A second evolution step has been connected to the integration of QKD systems into telecommunication networks. In a first stage, the effort concentrated in the development of backbone and metropolitan QKD network demonstrators to enable multi-user connectivity in Japan [8], the US [9], Europe [10], and China [11,12]. A nodal network of point-to-point (P2P) QKD links can then be used to relay a global key between any two distant locations in the network [8-11]. Alternative approaches based on active routing of optical signals were reported [12,13]. A second stage has focused on the integration of QKD systems into access networks. Here, a point-to-multiple-point (P2MP) architecture is more suitable to allow simultaneous access by multiple users rather than resorting to P2P QKD links. For instance, researchers at Toshiba demonstrated a quantum access network (QAN) of this type allowing a high-speed detector to be simultaneously used by up to 64 users [14]. This QAN was designed for resource and cost sharing where the most expensive components, the single photon detectors, were placed in the central location to be shared by multiple users.

The full capacity of QKD can be unleashed by incorporating the spatial division multiplexing (SDM) domain on top of the existing wavelength division multiplexing (WDM) layer. SDM is now being considered the route towards capacity upgrade for current core optical communication networks [15,16] and will progressively permeate into the access segment, especially if it has to be shared between passive optical networks (PONs) and Centralized Radio Access networks (CRANs) [17]. For QKD, SDM is especially attractive as multicore fibers (MCFs) can be designed to render a negligible inter-core crosstalk (< 50 dB) for distances of less than 50 km. Hence, they can be employed as a transmission media to increase the capacity of QKD systems, either by allowing the parallel transmission of independent keys between different final users sharing the same transmission medium (a feature desired in access networks) or by multiplying the capacity of key transmission by suitable disassembly, transmission and final assembly of a high-speed quantum key between two end users (a feature desired in backbone networks).

Several groups have reported experimental work connected to the use of MCFs for QKD transmission. In [18], researchers have demonstrated the successful transmission of QKD signal coexisting with classical data signals launched at full power in a 53-km 7-core fiber, while showing negligible degradation in performance. However, in this case, although the QKD signal was transmitted at a different core than the classical channels, it was coded in a different wavelength and thus the impact of MCF crosstalk could not be evaluated. More recent works have reported the transmission of multidimensional QKD channels in MCFs , however without including classical channels. In particular, orthogonal QKD channels where implemented by means of multiplexing of modes corresponding to mutually unbiased bases (MUBs). In [19], the operation of a four-dimensional QKD system encoded on the transverse modes of a 4-core MCF using a decoy state protocol has been demonstrated for a distance of 300 m. A more recent experiment [20] has reported a three-dimensional QKD system encoded using silicon photonic chips on the transverse modes provided by 4 cores of a 7-core MCF. The experimental results using a decoy state protocol have proved the successful transmission along up to 20 km.

The practical integration of QKD channels into optical fiber based SDM transmission media needs for a suitable system model that takes into account the transmission of both classical and quantum channels in order to evaluate the impact of each relevant physical parameter on the final performance of each channel and the feasibility of the design under consideration. To the best of our knowledge, this model has not been yet reported in the literature and it is the subject of this paper. Section 2 introduces the different SDM transmission media and system alternatives that we will consider in the paper. In particular, we consider homogeneous and heterogeneous MCFs as well as few-mode fibers for the transmission media, while for system configurations we consider multiple input multiple output (MIMO), single input multiple output (SIMO) and multiple input-single output (MISO) configurations with the possibility of arbitrarily choosing the number of classical and quantum channels transmitted over the SDM medium. Section 3 presents the models for the quantum channels. First, the Quantum Bit Error rate (QBER) expression is derived that takes into account the impact of intercore or intermodal crosstalk and the number of classical and quantum channels and, then, the expression of the net key rate is obtained both for standard BB84 and Decoy state protocol transmission under the assumption of photon number splitting (PNS) attack. Section 4 provides the derivation of the crosstalk coefficients for relevant designs of MCFs that will be employed in deriving the results for SDM-QKD transmission under realistic conditions in section 5. Finally, section 6 closes the paper with the relevant conclusions and future directions of research.

## 2.  Practical scenarios for SDM-QKD integration

Figure 1 presents the different SDM media and application scenarios that will be considered in this paper. For SDM media, we consider in first place homogeneous MCFs where all the cores are made of the same material and have the same geometry. Homogeneous MCFs are the most popular choice for capacity upgrade in SDM optical communication systems. Here we will assume that the MCF has $N$ different single-mode cores. Simple homogeneous MCFs can produce a considerable degree of intercore crosstalk if they are not properly designed and to reduce this effect it is customary either to modify the waveguide design by incorporating trenches or to resort to heterogeneous MCF designs, where the $N$ optical cores can feature different waveguide designs. Finally, although it will not be studied in detail in this paper, a few-mode fiber capable of transmitting $N$ orthogonal spatial modes can also be employed to integrate QKD with SDM.

The application scenarios are displayed in Figs. 1(b)-1(d). Figure 1(b) corresponds to a pure parallel or MIMO transmission system where the $N$ cores are employed to transmit $N$ spatially independent channels (one per core). The figure shows the case where all the channels are employed for QKD, but in general our model will consider the case where $N_q$ cores are employed to transmit $N_q$ quantum channels and $N-N_q$ are employed for classical transmission. The end points of each quantum channel $i = 1, 2,... N_q$ are implemented by means of an Alice transmitter #$i$ employing a faint pulse laser source (FPS) producing $\mu_i$ photons per bit at a repetition rate given by $f_{rep}$ that is phase modulated for key encoding (not shown in the figure) and by a Bob receiver #$i$ using an optical phase modulator for base selection (not shown in the figure) followed by single photon detector characterized by a responsivity $\rho_i$ and a dark count rate $d_{Bi}$. This configuration represents practical application scenarios ranging from parallel and simultaneous quantum and classical transmission in point to point links to capacity upgrade in a single transmitter/single receiver QKD system. In this an overall key rate $N$ times higher can be achieved by segmentation of the key in $N$ interleaved time blocks and sending each one using a different core.

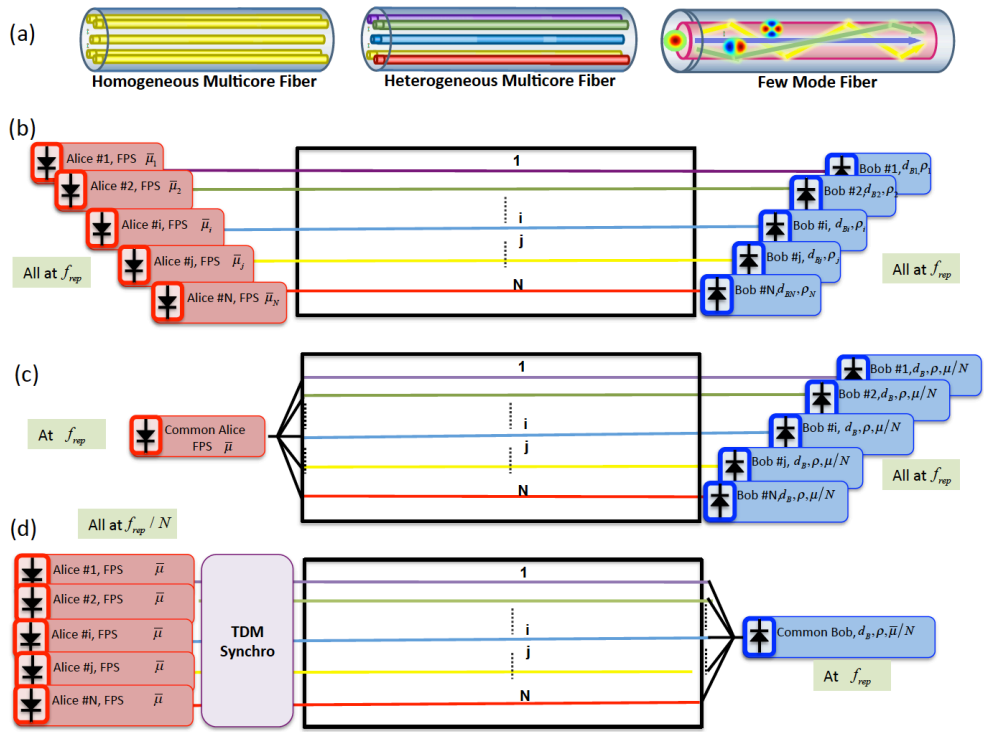

Fig. 1. Transmission media options (a) and application scenarios (b)-(d) for the integration of optical fiber QKD systems with Spatial Division Multiplexing.

Figure 1(c) corresponds to a SIMO transmission system representative of a PON downstream transmission from a central office to the end users in an access network. Here, a central office (i.e., a single Alice) distributes simultaneously several keys among different end users (Bob$_{\#1}$ to Bob$_{\#N}$). Alice transmitter uses a faint pulse laser source (FPS) producing $\mu$ photons per bit at a repetition rate given by $f_{rep}$ that is phase modulated for key encoding (not shown in the figure). The Bob receivers are assumed, for practical reasons to be identical (i.e. $\rho_i = \rho$ and $d_{Bi} = d_B$).

The last application scenario shown in Fig. 1(d) corresponds to a MISO transmission system representative of a PON upstream transmission from the end users to the central office in an access network. This scheme has been shown in [14] to be more efficient in terms of cost for multiple quantum key distribution in an access network as the high cost receiver is located at the central office and thus can be shared by the end-users. Here the $N$ end users (Alice #1 to Alice #N) independently agree on an individual key with the central office (i.e. single Bob). The Alice transmitters are assumed, for practical reasons to be identical (i.e. $\mu_i = \mu$) and operating at $f_{rep}/N$, as time division multiplexing (i.e. synchronization) is required to coordinate the medium access as in any ordinary upstream PON.

## 3. Model for Optical Fiber SDM-QKD transmission

### 3.1 Introductory remarks

The overall performance of a QKD transmission system is given by the Quantum Bit Error rate QBER. To derive a suitable QBER expression for SDM-QKD systems, we follow the approach of [21] specialized to an arbitrary core/mode. For SDM-QKD systems, we have to consider the influence of intercore/intermodal crosstalk terms depending on whether the transmission medium is a MCF or a few-mode fiber (FMF), respectively. For the derivation of the intercore/crosstalk term in a given core/mode due to photons leaking from other cores/modes, we follow a similar approach carried for other multiplexed QKD systems [22], where this term is added as an interference term. Once the QBER expression is presented, we will compute the net key rate, assuming decoy-state BB84 protocol [23] and PNS attacks. Expressions for other improved protocols can be certainly derived proceeding in a similar way, but they will not be in the scope of the paper.

### 3.2 QBER Expression

To derive a suitable QBER expression for SDM-QKD transmission in MCF/FMFs we follow the approach of [21] specialized to an arbitrary core/mode. At the detector placed to detect core/mode $n$ the probability that Bob detects a signal has three sources, one coming from the detection of signal photons, $p_{\exp}^{signal}(n)$, another coming from the detection of photons due to signal leak/crosstalk from each of the other cores/modes $p_{\exp}^{Xtalk}(m,n)$ and finally those arising from the dark counts of the specific detector $p_{\exp}^{d}(n)$. We will assume that, for each core/mode, the three sources are independent. Let $R_n(k)$ represent the probability that the source for mode/core $n$ sends $k$ photons then the probability that Bob's detector for mode/core $n$ is triggered by a signal photon can be expressed in terms of the detector efficiency $\rho_n$ and the end-to–end optical link transmission efficiency $T_{nn}$ as:

$$p_{\exp}^{signal}(n) = \sum_{k=1}^{\infty} R_n(k) \left[ \sum_{l=1}^{k} \binom{k}{l} (\rho_n T_{nn})^l (1 - \rho_n T_{nn})^{k-l} \right]. \quad (1)$$

In a similar way, the probability that Bob's detector for mode/core $n$ is triggered by a photon leaked from another core/mode $m$ can be expressed in terms of the detector efficiency $\rho_n$ and the end-to–end crosstalk coefficient (between core/mode $m$ and core/mode $n$) $XT_{mn}$ as:

$$p_{\exp}^{Xtalk}(m,n) = \sum_{k=1}^{\infty} R_m(k) \left[ \sum_{l=1}^{k} \binom{k}{l} (\rho_n XT_{mn})^l (1 - \rho_n XT_{mn})^{k-l} \right]. \quad (2)$$

For MCF/FMF transmission, the crosstalk coefficient $XT_{mn}$ is related to the fiber link length $L$ and the intercore/intermode average power coupling coefficient (PCC) $\bar{h}_{mn}$ (see Section 4 and [24]):

$$XT_{mn} = \tanh(\bar{h}_{mn} L) \approx \bar{h}_{mn} L. \tag{3}$$

Finally, the dark count distribution is simply given by:

$$p_{\exp}^{d}(n) = d_{B,n}. \tag{4}$$

In SDM-QKD systems, the signal source is provided de by laser pulses. In the case of a quantum channel it will be a strongly attenuated laser pulse, and therefore the photon number can be considered to be Poisson distributed with mean value $\bar{\mu}_n$. For core/modes carrying classical channels, the laser source will be represented by a coherent state with again, Poisson statistics, where the alpha parameter and the mean photon number per pulse are related by $\alpha = |\bar{\mu}_n|^2$ and $\bar{\mu}_n \gg 1$, [25]. For both cases then:

$$R_n(k) = \frac{e^{-\bar{\mu}_n}(\bar{\mu}_n)^k}{k!}. \tag{5}$$

Hence,

$$\begin{aligned}
p_{\exp}^{\text{signal}}(n) &= 1 - e^{-\rho_n T_{nn}\bar{\mu}_n} \approx \rho_n T_{nn}\bar{\mu}_n \\
p_{\exp}^{Xtalk}(m,n) &= 1 - e^{-\rho_n XT_{mn}\bar{\mu}_m} \approx \rho_n XT_{mn}\bar{\mu}_m = \rho_n \bar{h}_{mn} L \bar{\mu}_m \\
p_{\exp}^{Xtalk}(n) &= \sum_{\substack{m=1 \\ m\neq n}}^{N} p_{\exp}^{Xtalk}(m,n) = \sum_{\substack{m=1 \\ m\neq n}}^{N}\left(1 - e^{-\rho_n XT_{mn}\bar{\mu}_n}\right) \approx \sum_{\substack{m=1 \\ m\neq n}}^{N} \rho_n \bar{h}_{mn} L \bar{\mu}_m
\end{aligned} \tag{6}$$

where $p_{\exp}^{Xtalk}(n)$ is the overall crosstalk in core/mode $n$ due to the rest of interfering cores/modes.

The error rate stems from three sources again; the first is an error rate for the detected photons, which is due to alignment errors that impact over the interference visibility $V_n$ in a given core/mode $n$. In our case, it can be expressed as:

$$p_{\text{visibility}}^{Error}(n) = \frac{(1-V_n)}{2} p_{\exp}^{\text{signal}}(n) \approx \frac{(1-V_n)}{2}\rho_n T_{nn}\bar{\mu}_n. \tag{7}$$

The dark count contribution to the error probability is (only half of these photons contribute to errors):

$$p_{\text{dark}}^{error}(n) = \frac{p_{\exp}^{d}(n)}{2} = \frac{d_{B,n}}{2}. \tag{8}$$

Finally, the contribution of the intercore/intermode crosstalk signal to the error is:

$$p_{Xtalk}^{Error}(n) = \frac{p_{\exp}^{Xtalk}(n)}{2} \approx \sum_{\substack{m=1 \\ m\neq n}}^{N}\frac{\rho_n \bar{h}_{mn} L \bar{\mu}_m}{2}. \tag{9}$$

From (1)-(9) we can obtain an expression for the QBER in a given core/mode $n$:

$$QBER(n) = Q_n = \frac{\text{Error Photons received by all contributions}}{\text{Photons received by all contributions}} = \qquad (10)$$

$$= \frac{p_{visibility}^{Error}(n) + p_{dark}^{Error}(n) + p_{Xtalk}^{Error}(n)}{p_{exp}^{signal}(n) + p_{exp}^{Xtalk}(n) + p_{exp}^{d}(n)} \approx \frac{\frac{(1-V_n)}{2}\rho_n T_{nn}\bar{\mu}_n + \frac{d_{B,n}}{2} + \sum_{\substack{m=1\\m\neq n}}^{N} \frac{\rho_n \bar{h}_{mn} L \bar{\mu}_m}{2}}{\rho_n T_{nn}\bar{\mu}_n + d_{B,n} + \sum_{\substack{m=1\\m\neq n}}^{N} \rho_n \bar{h}_{mn} L \bar{\mu}_m},$$

which takes into account all the main relevant physical parameters.

### 3.3 Net key Rates

To calculate the net key rate we assume that the quantum channels use a Decoy-State BB84 protocol and are subject to photon number splitting attack (PNS). More elaborate/efficient protocols [26] can be also considered but the main results/conclusion will not change substantially. Regarding the impact of classical channels, we will assume that $\bar{\mu}_n = \bar{\mu}_c \gg 1$. The net key rate for channel $n$ is given by [2]:

$$R_{net,n} = \frac{1}{2}\rho_n T_{nn}\bar{\mu}_{opt,n} f_{rep}\left[1 - 2h(Q_n)\right], \qquad (11)$$

where the optimum value for the number of photons per bit is given by:

$$\bar{\mu}_{opt,n} \approx \frac{1}{2}\left[1 - \frac{h(Q_n)}{1-h(Q_n)}\right], \qquad (12)$$

$$h(Q_n) = -Q_n \log_2(Q_n) - (1-Q_n)\log_2(1-Q_n).$$

Note that $\bar{\mu}_{opt,n}$ is both in the $R_{net,n}$ and $Q_n$ so we have to solve first an implicit equation to get $Q_n$. If all the transmitted channels in the SDM system are quantum then:

$$Q_n = \frac{\frac{(1-V_n)}{2}\rho_n T_{nn}\frac{1}{2}\left[1-\frac{h(Q_n)}{1-h(Q_n)}\right] + \frac{d_{B,n}}{2} + \sum_{\substack{m=1\\m\neq n}}^{N} \frac{\rho_n \bar{h}_{mn} L}{4}\left[1-\frac{h(Q_m)}{1-h(Q_m)}\right]}{\rho_n T_{nn}\frac{1}{2}\left[1-\frac{h(Q_n)}{1-h(Q_n)}\right] + d_{B,n} + \sum_{\substack{m=1\\m\neq n}}^{N} \frac{\rho_n \bar{h}_{mn} L}{2}\left[1-\frac{h(Q_m)}{1-h(Q_m)}\right]}. \qquad (13)$$

Equation (13) for $n = 1, 2, \ldots N$ defines a system of $N$ implicit equations rendering the values of $Q_n$. From these values, one can get $\bar{\mu}_{opt,n}$ and the net key rate using Eqs. (12) and (11), respectively.

If $N_q$ out of the $N$ SDM channels are quantum then:

$$Q_n = \frac{\frac{(1-V_n)}{2}\rho_n T_{nn}\frac{1}{2}\left[1-\frac{h(Q_n)}{1-h(Q_n)}\right] + \frac{d_{B,n}}{2} + \sum_{\substack{m=1\\m\neq n}}^{N_q} \frac{\rho_n \bar{h}_{mn} L}{4}\left[1-\frac{h(Q_m)}{1-h(Q_m)}\right] + \sum_{\substack{m=1\\m\neq n}}^{N-N_q} \frac{\rho_n \bar{h}_{mn} L \bar{\mu}_c}{2}}{\rho_n T_{nn}\frac{1}{2}\left[1-\frac{h(Q_n)}{1-h(Q_n)}\right] + d_{B,n} + \sum_{\substack{m=1\\m\neq n}}^{N_q} \frac{\rho_n \bar{h}_{mn} L}{2}\left[1-\frac{h(Q_m)}{1-h(Q_m)}\right] + \sum_{\substack{m=1\\m\neq n}}^{N-N_q} \rho_n \bar{h}_{mn} L \bar{\mu}_c}, \qquad (14)$$

where the $m$ subindex runs along the independent ordered lists of quantum and classical channels in each case respectively. Equation (14) for $n = 1, 2, \ldots N_q$ defines a system of $N_q$ implicit equations rendering the values of $Q_n$. From these values, one can get $\bar{\mu}_{opt,n}$ and the net key rate using Eqs. (12) and (11), respectively.

## 4. Multicore fiber crosstalk performance

### 4.1. Average power coupling coefficients

We model the intercore crosstalk following the analytical expressions derived in [27] by Koshiba et al. for the average power coupling coefficient (PCC) $h_{mn}$ between cores $m$ and $n$. If we consider the multicore fiber is bent at a constant bending radius $R_b$ while being twisted at a constant rate $\gamma$, then the longitudinal varying local PCC, $h_{mn}(z)$, can be described accurately by assuming an exponential autocorrelation function. Then, we can obtain the analytical expression of the average PCC by averaging over a twist pitch $2\pi/\gamma$ as:

$$\bar{h}_{mn} = \sqrt{2}\kappa_{mn}^2 d \left| \frac{1}{\sqrt{a(b+\sqrt{ac})}} + \frac{1}{\sqrt{c(b+\sqrt{ac})}} \right|, \quad (15)$$

where $\kappa_{mn}$ is the redefined mode-coupling coefficient, which has been computed from (Eq. (33), [28]), $d$ is the correlation length and

$$\begin{aligned}
a &= 1 + \left(\Delta\beta_{mn}d - \frac{B_{mn}d}{R_b}\right)^2, \\
b &= 1 + \left(\Delta\beta_{mn}d\right)^2 - \left(\frac{B_{mn}d}{R_b}\right)^2, \\
c &= 1 + \left(\Delta\beta_{mn}d + \frac{B_{mn}d}{R_b}\right)^2, \\
B_{mn} &= \sqrt{(\beta_m x_m - \beta_n x_n)^2 + (\beta_m y_m - \beta_n y_n)^2},
\end{aligned} \quad (16)$$

being $\Delta\beta = \beta_m - \beta_n$ and $x_m$, $y_m$ the coordinates of the center of core $m$ at $z = 0$. The analytical expression in Eq. (15) accounts for a generic MCF, either with homogeneous or heterogeneous cores, including as well the case of trench-assisted structures. From, this coefficient, one can easily derive the crosstalk between cores $m$ and $n$ a for an $L$-km fiber link as:

$$XT = \tanh(\bar{h}_{mn}L), \quad (17)$$

which can be approximated as $\bar{h}_{mn}L$ for low values of $\bar{h}_{mn}$.

### 4.2. Homogeneous Multicore fiber

Different levels of intercore crosstalk are considered by evaluating two representative homogeneous MCFs that are characterized by different PCCs. The first one is a 7-core fiber analyzed by Cartaxo and Alves in [29], which is characterized by a core pitch $\Lambda$ of 30 μm, a cladding diameter CD of 125 μm, a core radius $a$ of 4 μm and a relative refractive index difference between core and cladding of $\Delta_1 = 0.5\%$. The second MCF provides a lower level of crosstalk mainly because the core size is reduced significantly. It is a commercially available 7-core fiber provided by Fibercore (SM-7C1500), [30]. It has a core pitch $\Lambda$ of 35 μm, a cladding diameter CD of 125 μm, an estimated core radius $a = 3$ μm and an estimated relative index difference between core and cladding $\Delta_1 = 0.96\%$ (estimated from the parameters given in its specifications: numerical aperture in the range [0.20-0.22] and mode field diameter in the range [5.7-6.5] μm at an optical wavelength of 1550 nm). Figure 2 shows the dependence of the average PCC $\bar{h}_{mn}$ with the bending radius, which was calculated from Eq. (15), for both homogeneous MCFs. As expected, the PCC increases linearly with the bending radius due to the reduction in the index difference between the cores, reaching its theoretical maximum value if the fiber were totally straight. The significant difference between both PCC responses arises from the fact that the Fibercore fiber has a higher core pitch and better mode confinement.

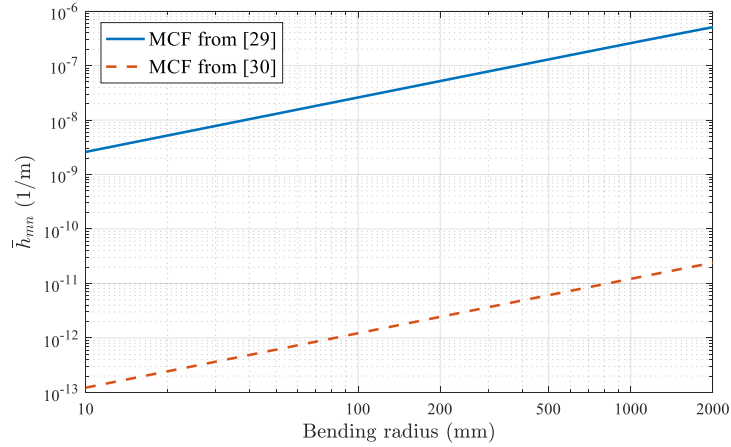

Fig. 2. Average power coupling coefficient dependence on the bending radius for the evaluated homogenenous MCFs.

### 4.3. Heterogeneous multicore fiber

Even though the second homogeneous MCF evaluated previously provided a good crosstalk performance for classical data transmission, the range of computed average PCC values care still be too high for certain scenarios of combined QKD and classical channel transmission as it will be shown in section 5. With the aim of reducing the intercore crosstalk furtherly, we designed a particular ultralow-crosstalk 7-core MCF built upon trench-assisted heterogeneous cores. The design is based on three different types of cores in terms of core radius and dopant concentration, where one type of core is used as central core and the other two types as outer cores so neighbouring cores are dissimilar. Figure 3 (a) depicts the cross-section of the designed fiber with its three types of cores (type A: blue color, type B: yellow color and type C: orange color) and Fig. 3 (b) shows the refractive index profile of a trench-assisted core.

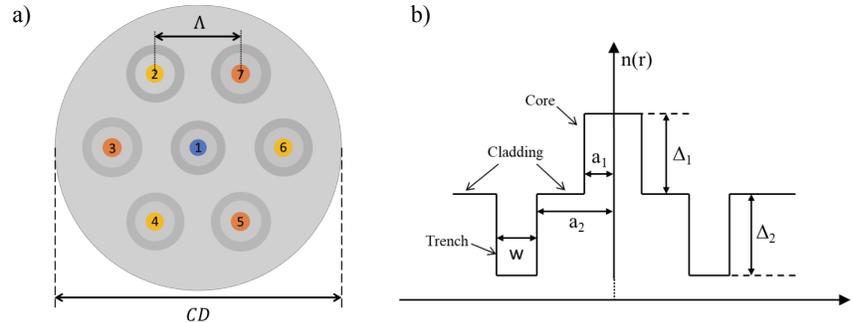

Fig. 3. (a) Cross-section of the designed heterogeneous trench-assisted MCF (each color illustrates a different type of core); (b) Core refractive index profile, being $a_1$: core radius; $a_2$: inner cladding radius; $w$: trench width; $\Delta_1$: core-to-cladding relative index difference; $\Delta_2$: cladding-to-trench relative index difference.

The design strategy we followed supports three main goals. First, the average PCC should be at least as low as $10^{-14}$, as we demonstrate in in section 5, to allow both pure quantum and hybrid quantum/classic transmission without significant penalties. Second, as agreed in [27], we must assure a critical bending radius lower than 10 cm to avoid an undesired crosstalk increase due to phase matching between cores when the fiber is bent. Third, the cores must remain in the singlemode operation regime as to avoid intermodal crosstalk.

The parameters of the heterogeneous MCF design and the propagation characteristics (effective index $n_{eff}$ and effective area $A_{eff}$) computed with the Fimmwave software from Photon Design are gathered in Table 1. To further reduce the average crosstalk, we set the core pitch to 45 μm and the cladding diameter to 150 μm.

Table 1. Parameters of the heterogeneous MCF designed

| Parameter | Core A | Core B | Core C | Units |
|---|---|---|---|---|
| $a_1$ | 4.2 | 4.5 | 4.7 | μm |
| $a_2/a_1$ | | 2.0 | | - |
| $w/a_1$ | | 1.0 | | - |
| $\Delta_1$ | 0.58% | 0.48% | 0.38% | - |
| $\Delta_2$ | | -0.63% | | - |
| $n_{cl}$ | | 1.444 | | - |
| $n_{eff}$ | 1.449 | 1.448 | 1.447 | - |
| $A_{eff}$ | 60.31 | 70.55 | 80.91 | μm² |
| CD | | 150 | | μm |
| Λ | | 45 | | μm |

Figure 4 shows the average PCC $\bar{h}_{mn}$ for any possible core pair combination. As characteristic of heterogeneous MCFs, we observe a peak in the magnitude of PCC caused by forced phase matching between cores for a specific value of the bending radius that is called the critical bending radius. Although this particular design aimed to reduce the magnitude of this crosstalk peak as much as possible, we must bear in mind that the optimum working zone is the one corresponding to bending radii above this critical value. For the core combinations represented here, one critical bending radius is located at 33 mm while the other two at 65 mm (worst case). Total crosstalk can be calculated using Eq. (17) for a given fiber length. For example, a length of 1 km and an average PCC value of $10^{-14}$, result in a crosstalk level of -110 dB.

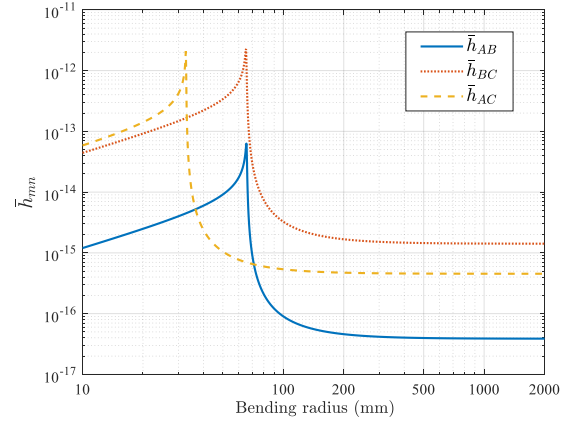

Fig. 4. Average power coupling coefficient dependence on the bending radius for different core pair combinations of the designed heterogeneous MCF. The blue solid line is the average PCC between type A and B cores, red dotted line is the average PCC between type B and C cores and yellow dashed line is the average PCC between type A and C cores.

## 5. Results and Discussion

### 5.1 Introduction

We proceed now to analyze the possibility of multiple channel QKD transmission using MCFs. For this purpose, we employ the model developed in section 3 together with the relevant parameters of representative MCFs given in section 4. In addition, we consider typical values for the parameters characterizing the pulse sources and photon counters given in table 2. We will assume that these parameters are the same for these devices regardless the core number, that is: $d_{B,n} = d_B$, $\rho_n = \rho$, $V_n = V$, $T_{nn} = T$, $f_{rep,n} = f_{rep}$.

Table 2. Faint pulse/Classical laser source, photon counter and fiber loss typical parameter values employed in the simulations

| Simulation parameter | Physical concept | Value/Range |
|---|---|---|
| $d_B$ | Photon counter dark count rate | $10^{-5}$ |
| $\rho$ | Photon counter responsivity | 0.1 (A/W) |
| $f_{rep}$ | Pulse repetition rate at transmitter | 10 MHz |
| $V$ | Receiver visibility | 0.98 |
| $\mu_c$ | Equivalent number of photons per pulse in the classical channels | $10^6$ |
| $\mu_{opt}$ | Optimum number of photons per pulse in quantum channels | 0.1-0.5 as given by the application of Eqs. (12)-(13) |
| $T = 10^{-\frac{\alpha}{10}L}$ $\alpha = 0.2$ dB/km | Transmission factor core n, length L | Depending on the link length |

### 5.2 Quantum channel transmission only

We consider in first place the case where only quantum channels are transmitted through the SDM optical fiber. For practical reasons, we consider a 7-core homogeneous MCF as the transmission medium as it is the most straightforward commercially available solution. We consider two cases of MCFs representative of moderate and low inter-core crosstalk levels that have been described in section 4. In the first case, the maximum values of the inter-core coupling coefficient are around $h = 10^{-7}$ m$^{-1}$, while in the second they are around $h = 10^{-11}$ m$^{-1}$. Figure 5 shows the obtained results for the capacity and optimum average photon number as a function of the MCF link length after solving Eqs. (10)-(12). Figures 5(a) and 5(b) show the results for the moderate intercore crosstalk MCF. As it can be observed, the outer peripheral cores show an identical performance as expected as each one is subject mainly to the crosstalk of three closest neighboring cores. The central core displays worse results as it is subject to the crosstalk of 6 closest neighboring cores, hence the capacity is $R_{net,central\_core} + 6R_{net,i}$, where $i$ stands for any peripheral core. If, however, the inter-core crosstalk is low, then its effect is almost negligible, as shown in Figs. 5(c) and 5(d), where it is clear that all cores feature an identical performance in terms of net capacity and present equal values for the optimum number of photons per bit. Note as well that in this second case the net rate is $7R_{net,i}$, and $i$ stands for any core.

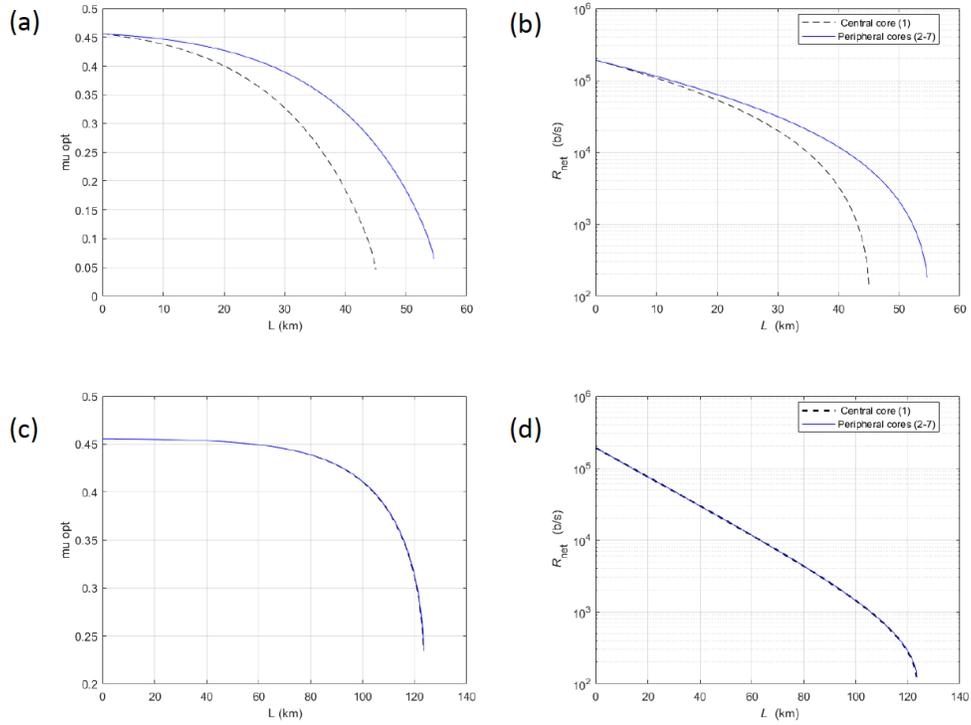

Fig. 5. (a) Optimum number of photons per bit vs link distance $L$ in km and (b) Net quantum transmission rate per core vs link distance $L$ in km for a moderate intercore crosstalk (the coupling coefficients are in the order of $h = 10^{-7}$ m$^{-1}$) 7- core homogeneous MCF. (c) Optimun number of photons per bit vs link distance $L$ in km and (d) Net quantum transmission rate per core vs link distance $L$ in km for a low intercore crosstalk ($h = 10^{-11}$ m$^{-1}$) 7-core homogeneous MCF. Simulation parameters are given in Table 2.

The results indicate that, in principle, multiple quantum channel transmission is viable both for moderate as well as for low inter-core crosstalk 7-core homogeneous MCFs. In the first case, at least 6 cores are expected to display identical performance in terms of net key rate and distances of up to 40 km are reachable with key rates ranging from 3 to 10 kb/s per core, which covers the access segment distance range. In the second, this figure can eventually rise to identical behavior of the 7 cores and transmission distances around 100 km for key rates over 1 kb/s per core. This covers access as well as moderate distance metropolitan segments.

*5.3 Simultaneous transmission of Quantum and Classical channels*

We start by considering the case of introducing one classical channel in the SDM medium. In this case, the classical transmitter is a pulsed optical transmitter providing standard output power values (i.e., between 1 and 10 mW). This is well represented by a coherent state with $\mu_c = 10^6$. The rest of the cores transmit quantum channels. Two cases of interest are identified that correspond to transmitting the classical channel through the central core or a peripheral one, respectively.

Figure 6 plots the obtained results for these two cases when the SDM transmission medium is a low intercore crosstalk ($h = 10^{-11}$ m$^{-1}$) 7-core MCF. Figure 6(a) givers the results for the case where the classical core is sent through the central core, while Fig. 6(b) depicts the same results when the classical channel is transmitted by a peripheral core.

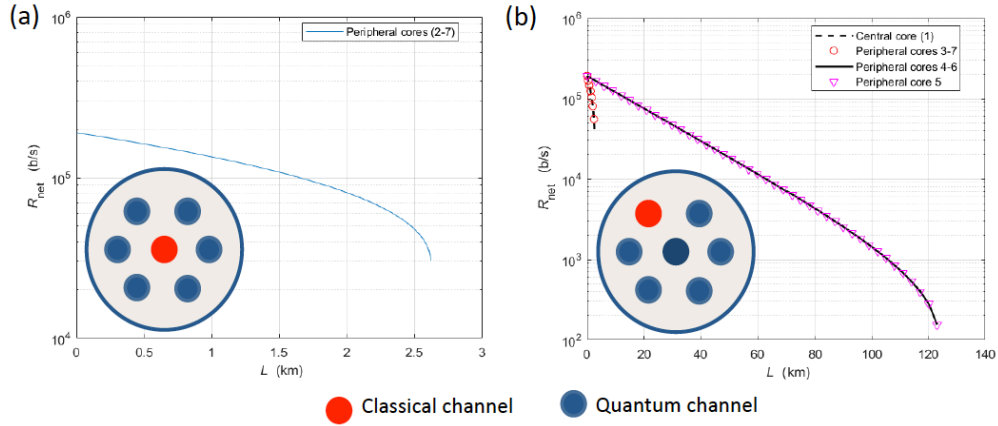

Fig. 6. Net key rate for QKD channels versus the link distance in an SDM-QKD link supporting one classical channel and 6 QKD channels and implemented using a 7-core homogeneous MCF with intercore coupling coefficients around $h = 10^{-11}$ m$^{-1}$. (a) Case where the classical channel is transmitted by the central core. (b) Case where the classical channel is transmitted by peripheral core #2.

As it can be appreciated, the introduction of a classical channel in the central core has a profound impact over the maximum reach of the rest of quantum channels, posing a severe restriction, in the range of 2-2.5 km. In practical terms, this makes QKD transmission unfeasible unless a very short reach distance is targeted. When the classical channel is transmitted through a peripheral core, it impacts the central core and the other two near neighbors in the peripheral area, however, the impact over the other three peripheral cores is almost negligible. Yet in this last case, though QKD transmission is feasible in three cores, other three cores must be left unused, as they can transmit neither quantum channels (they would be spoiled by classical interference) nor classical channels, as they would spoil the QKD transmission in the cores where it is feasible.

To accommodate both quantum and classical channels and be able to use all the cores in the MCFs, further simulations have shown that one needs values for the coupling coefficient in the range of $h = 10^{-14}$ m$^{-1}$ or smaller. These can only be attained by means of heterogeneous MCFs and, in particular, we consider the ultralow-crosstalk design reported in section 4.

Figure 7, shows the results for different cases of combined classical and quantum channel transmission. Intercore coupling coefficients between the three groups of cores are, as shown in section 4, $h_{1,2} = 4\times10^{-17}$, $h_{2,3} = 4\times10^{-16}$ and $h_{1,3} = 1.5\times10^{-15}$. Figure 7 (a) corresponds to the case where a single classical channel is transmitted through the central core (core #1) and quantum channels are delivered in the rest of the peripheral cores (#2 to #7). As it can be appreciated, both transmission regimes are supported by the MCF. Key rates for QKD channels are almost identical within the linear range (up to 80 km), with a departure from that attainable in cores belonging to the group formed by #2, #4, #6 and that attainable in cores belonging to the group formed by #3, #5, #7. This behavior is to be expected as the inter-core coupling coefficients with core #1 are different. Since cores #2, #4, #6 have a very small (in fact negligible) coupling coefficient with #1, they feature a performance in terms of net key rate identical to a channel with zero coupling as shown in the inset of Fig. 7(b). Channels #3, #5, #7 have a slightly higher coupling coefficient ($h_{1,3}$), which is enough to result in a different (i.e. worse) key rate for higher distances. In any case, it is clear that both the classical and quantum channels can coexist within the heterogeneous MCF transmission link.

Figure 7 (b) corresponds to the case where a single classical channel is transmitted through a peripheral core (core #2). In this case, QKD transmission is feasible in the rest of the cores with the same key rates, which corresponds to negligible crosstalk impact as it can be appreciated by comparing to the inset of Fig. 7(b). Equal performance can be attained up to 100 km and key rates of 1 kb/s per core. The possibility of using more than one core for classical transmission is demonstrated by the results given in Figs. 7(c) and 7(d).

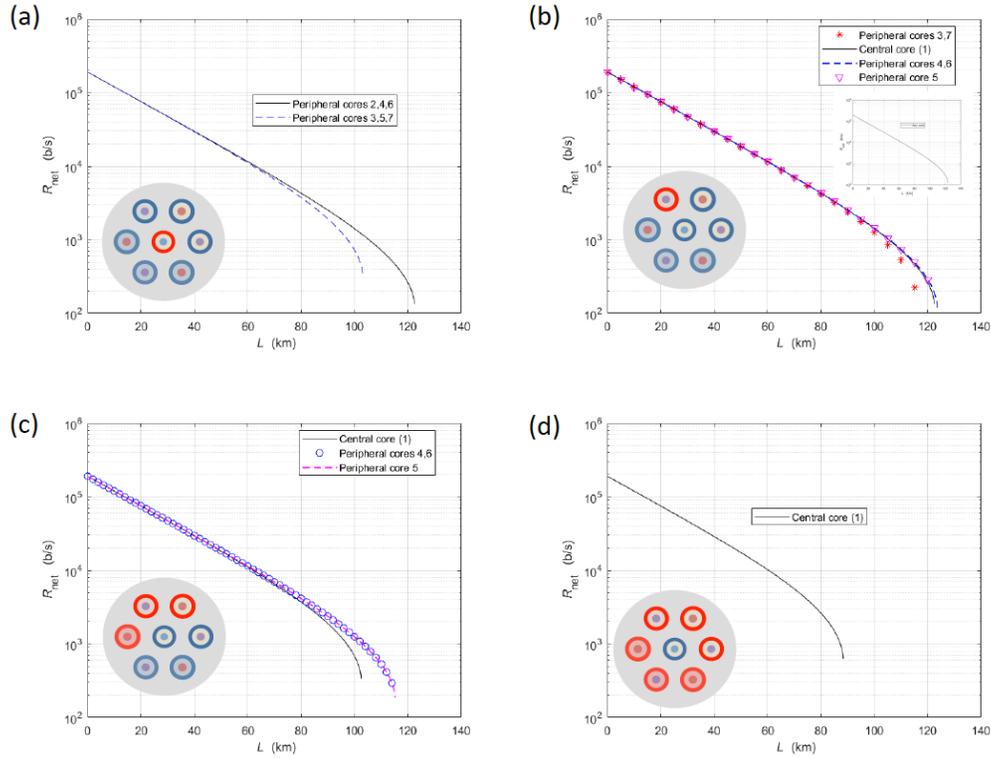

Fig. 7. Net quantum transmission rate per core vs link distance $L$ in km for very low inter-core crosstalk (coupling coefficients range from $h = 10^{-15}$ m$^{-1}$ to $h = 10^{-17}$ m$^{-1}$) 7-core heterogeneous MCF. (a) One classical channel transmitted by the central core (core #1) and 6 QKD channels (cores #2-#7). (b) One classical channel transmitted by the peripheral core (core #2) and 6 QKD channels (cores #1 and #3-#7). (c) Three classical channels transmitted by cores #2, #3 and #7 and four QKD channels (#1, #4, #5 and #6). (d) One QKD channel transmitted by the central core (#1) and six classical channels transmitted by the peripheral cores (#2-#7). Simulation parameters are given in Table 2. Color code for cores transmitting classical and quantum channels is the same as in Fig. 6.

Figure 7(c) illustrates the results for the case where 3 cores (#2, #3 and #7) are employed for classical transmission and 4 (#1, #4, #5 and #6) are employed for QKD. Again, QKD transmission with almost equal net key rate per core is feasible, with a slight worse value for the central core (#1) that experiences the highest inter-core crosstalk values. Equal core performance up to around 80 km and key rates of 50 kb/s per core are possible. Finally, Fig. 7(d) shows the case where 6 classical channels and one QKD (in core #1) are transmitted. This corresponds to the highest impact of intercore crosstalk. Again, in this case QKD transmission is feasible up to distances around 80 km and key rates in the range of 30-40 kb/s. These results should be compared to those of Fig. 6.

*5.4 Discussion*

The results obtained in sections 5.3 and 5.4 confirm that homogeneous MCFs can accommodate the transmission of spatially multiplexed quantum channels. However and even though inter-core crosstalk levels in these fibers are small and even negligible for classical applications, they are too high for systems where classical and quantum channels need to be multiplexed. In these cases, the inter-core crosstalk must be further reduced and this requires the use of heterogeneous MCFs.

The model developed in section 3 can be applied to other types of MCFs and QKD protocols. The procedure should be similar to the one described in sections 4 and 5. First, values of pertinent intercore/intermode coupling coefficients are required and this depends on the particular optical fiber transmission medium under consideration. In the case of MCFs, the extension to 19-core and 36-core fibers should take into account the fact that different cores are surrounded by different numbers of near neighbors, thus more than the three values of $h$ coefficients will be obtained. This is not a limitation as Eqs. (10)-(14) are derived to take this into account. As related to QKD protocols, the model can be extended to more powerful versions, such as the T12 [26]. In this case, Eqs. (10)-(14) must be replaced by those giving QBER and net key rate of the protocol. It can also be applied to systems aiming at combining SDM with other types of multiplexing, such as Wavelength Division (WDM) or Subcarrier (SCM) multiplexing. In this case, the model must be upgraded to include the contributions of other wavelengths and/or subcarriers, following the procedure described in the literature [31-34].

## 6. Summary and Conclusions

We have reported a model to evaluate the performance of multiple QKD channel transmission using spatial division multiplexing in multicore and few-mode fibers. The model has been then employed to analyze the feasibility of QKD transmission in 7-core MCFs for two scenarios of practical interest. The first accounts for the use of SDM to transmit only QKD channels. Here, we have found that standard homogeneous MCFs enable transmission distances per core compatible with transmission parameters (distance and net key rate) very close to those of single core singlemode fibers and, therefore, parallel QKD transmission of $N$ channels or a x$N$ multiplication of a single channel key rate over a given link distance is possible. The second scenario accounts for the simultaneous transmission of QKD and classical channels. Here we found that homogeneous MCFs cannot support this transmission regime as the inter-core crosstalk is too high. Heterogeneous MCFs can be designed to reduce this value to such levels that make this option feasible. Future work should address the extension of the model to MCFs with a higher core count (19, 36) and more efficient QKD protocols.

*Acknowledgments*

European Regional Development Fund (ERDF); Galician Regional Government (project GRC2015/018 and agreement for funding AtlantTIC (Atlantic Research Center for Information and Communication Technologies)); European Research Council (ERC)

(Consolidator Grant 724663); Spanish MINECO (TEC2016-80150-R project and Ramon y Cajal fellowship RYC-2014-16247 for I. Gasulla).

## References

1. N. Gisin, G. Ribordy, W. Tittel, and H. Zbinden, "Quantum cryptography," Rev. Mod. Phys. **74**, 145–195 (2002).
2. V. Scarani, H. Bechmann-Pasquinucci, N. J. Cerf, M. Dušek, N. Lütkenhaus, and M. Peev, "The security of practical quantum key distribution," Rev. Mod. Phys. **81**(3), 1301–1350 (2009).
3. A. Ekert and R. Renner, Nature **507**, 443 (2014).
4. M. Tomamichel, C. C. W. Lim, N. Gisin, and R. Renner, "Tight finite-key analysis for quantum cryptography," Nat. Commun. **3**, 634 (2012).
5. D. Gottesman, H. K. Lo, N. Lütkenhaus, and J. Preskill, "Security of quantum key distribution with imperfect devices," Quantum Information & Computation **4**, 325 (2004).
6. M. Lucamarini, K. A. Patel, J. F. Dynes, B. Fröhlich, A. W. Sharpe, A. R. Dixon, Z. L. Yuan, R. V. Penty and A. J. Shields, "Efficient decoy-state quantum key distribution with quantified security," Opt. Express **21**(21), 24550 (2013).
7. Z. L. Yuan, B. E. Kardynal, A. W. Sharpe, and A. J. Shields, "High speed single photon detection in the near infrared," Appl. Phys. Lett. **91**, 041114 (2007).
8. M. Sasaki et al., "Field test of quantum key distribution in the Tokyo QKD Network," Opt. Express **19**(11), 10387 (2011).
9. C. Elliot, A. Colvin, D. Pearson, O. Pikalo, J. Schlafer, and H. Yeh, "Current status of the DARPA quantum network," arXiv:quant-ph/0503058.
10. M. Peev, C. Pacher, R. Alleaume, C. Barreiro, J. Bouda, W. Boxleitner, T. Debuisschert, E. Diamanti, M. Dianati, J. F. Dynes et al., "The SECOQC quantum key distribution network in Vienna," New J. Phys. **11**, 075001 (2009).
11. T.-Y. Chen, J. Wang, H. Liang, W.-Y. Liu, Y. Liu, X. Jiang, Y. Wang, X. Wan, W.-Q. Cai, L. Ju, L.-K. Chen, L.-J. Wang, Y. Gao, K. Chen, C.-Z. Peng, Z.-B. Chen, and J.-W. Pan, "Metropolitan all-pass and inter-city quantum communication network," Opt. Express **18**(26), 27217-27225 (2010).
12. S. Wang et al., "Field-test of wavelength-saving quantum key distribution network," Opt. Lett. **35**, 2454 (2010).
13. A. Ciurana, J. Martínez-Mateo, M. Peev, A. Poppe, N. Walenta, H. Zbinden, and V. Martín, "Quantum metropolitan optical network based on wavelength division multiplexing," Opt. Express **22**(2), 1576-1593 (2014).
14. B. Fröhlich, J. F. Dynes, M. Lucamarini, A. W. Sharpe, Z. Yuan, and A. J. Shields, "A quantum access network," Nature **501**, 69-72 (2013).
15. P. J. Winzer, D. T. Neilson, and A. R. Chraplyvy, "Fiber-optic transmission and networking: the previous 20 and the next 20 years [Invited]," Opt. Express **26**(18), 24190-24239 (2018).
16. B. Shariati, A. Mastropaolo, N-P. Diamantopoulos, J. M. Rivas-Moscoso, D. Klonidis, and I. Tomkos, "Physical-Layer-Aware Performance Evaluation of SDM Networks Based on SMF Bundles, MCFs, and FMFs," J. Opt. Commun. Netw. **10**, 712-722 (2018).
17. J. Galvé, I. Gasulla, S. Sales and J. Capmany, "Reconfigurable radio access networks using multicore fibers," J. Quantum Electr. **52**, 0600507 (2016).
18. J. F. Dynes, S. J. Kindness, S. W.-B. Tam, A. Plews, A. W. Sharpe, M. Lucamarini, B. Fröhlich, Z. L. Yuan, R. V. Penty, and A. J. Shields, "Quantum key distribution over multicore fiber," Opt. Express **24**(8), 8081-8087 (2016).
19. G. Cañas, N. Vera, J. Cariñe, P. González, J. Cardenas, P. W. R. Connolly, A. Przysiezna, E. S. Gómez, M. Figueroa, G. Vallone, P. Villoresi, T. Ferreira da Silva, G. B. Xavier, and G. Lima, "High-dimensional decoy-state quantum key distribution over multicore telecommunication fibers," Phys. Rev. A **96**, 022317 (2017).
20. Y. Ding, D. Bacco, K. Dalgaard, X. Cai, X. Zhou, K. Rottwitt, and L. K. Oxenløwe, "High-dimensional quantum key distribution based on multicore fiber using silicon photonic integrated circuits," npj Quantum Information **3,** 25 (2017).
21. N. Lütkenhaus, "Security against individual attacks for realistic quantum key distribution," Phys. Rev. A, (052304) (2000).
22. J. Capmany, "Photon nonlinear mixing in subcarrier multiplexed quantum key distribution systems," Optics Express **17**(8), 6457–6464 (2009).
23. H. K. Lo, X. Ma and K. Chen, "Decoy state quantum key distribution," Phys. Rev. Lett. **94,** 230504 (2005).
24. M. Koshiba, K. Saitoh, K. Takenaga, and S. Matsuo, "Analytical Expression of Average Power-Coupling Coefficients for Estimating Intercore Crosstalk in Multicore Fibers," IEEE Photonics Journal **4**(5), 1987-1995 (2012).
25. C. G. Gerry and P. L. Kight, Introductory Quantum Optics, Cambride Unievrsity Press, (2005).
26. M. Lucamarini, K. A. Patel, J. F. Dynes, B. Fröhlich, A. W. Sharpe, A. R. Dixon, Z. L. Yuan, R. V. Penty, and A. J. Shields, "Efficient decoy-state quantum key distribution with quantified security," Opt. Express **21**(21), 24550-24565 (2013).
27. M. Koshiba, K. Saitoh, K. Takenaga, and S. Matsuo, "Analytical Expression of Average Power-Coupling Coefficients for Estimating Intercore Crosstalk in Multicore Fibers," IEEE Photonics J. **4**(5), 1987-1995 (2012).
28. J. Tu, K. Saitoh, M. Koshiba, K. Takenaga, and S. Matsuo, "Design and analysis of large-effective-area heterogeneous trench-assisted multi-core fiber," Opt. Express **20**(14), 15157-15170 (2012).
29. T. Cartaxo and F. Alves, "Discrete Changes Model of Inter-core Crosstalk of Real Homogeneous Multi-core Fibers," J. Lightwave Technol. **35**(12), 2398-2408 (2017).

30. Fibercore, "Multicore fiber datasheet", available at https://www.fibercore.com/mediaLibrary/images/english/8938.pdf, (2016).
31. I. Choi, R. J. Young, and P. D. Townsend, "Quantum key distribution on a 10Gb/s WDM-PON," Opt. Express **18**(9), 9600-9612 (2010).
32. J. Mora, W. Amaya, A. Ruiz-Alba, A. Martinez, D. Calvo, V. García Muñoz, and J. Capmany, "Simultaneous transmission of 20x2 WDM/SCM-QKD and 4 bidirectional classical channels over a PON," Opt. Express **20**(15), 16358-16365 (2012).
33. J. Mora, A. Ruiz-Alba, W. Amaya, A. Martínez, V. García-Muñoz, D. Calvo, and J. Capmany, "Experimental demonstration of subcarrier multiplexed quantum key distribution system," Opt. Lett. **37**(11), 2031-2033 (2012).
34. A. V. Gleim, V. I. Egorov, Yu. V. Nazarov, S. V. Smirnov, V. V. Chistyakov, O. I. Bannik, A. A. Anisimov, S. M. Kynev, A. E. Ivanova, R. J. Collins, S. A. Kozlov, and G. S. Buller, "Secure polarization-independent subcarrier quantum key distribution in optical fiber channel using BB84 protocol with a strong reference," Opt. Express **24**(3), 2619-2633 (2016).